\documentclass[12pt]{article}

\usepackage{graphicx}
\usepackage{dsfont}
\usepackage{amssymb}
\usepackage{amsmath}
\usepackage{mathrsfs}

\begin{document}
\title{General Relativity in Electrical Engineering}
\author{Ulf Leonhardt and Thomas G. Philbin\\
School of Physics and Astronomy, University of St Andrews,\\
North Haugh, St Andrews KY16 9SS, Scotland
}
\date{\today}
\maketitle
\begin{abstract}
In electrical engineering
metamaterials have been developed 
that offer unprecedented control over
electromagnetic fields.
Here we show that general relativity lends
the theoretical tools for designing devices 
made of such versatile materials.
Given a desired device function,
the theory describes the electromagnetic properties
that turn this function into fact.
We consider media that facilitate
space-time transformations
and include negative refraction.
Our theory unifies the concepts 
operating behind the scenes of
perfect invisibility devices,
perfect lenses,
the optical Aharonov-Bohm effect
and electromagnetic analogs of the event horizon,
and may lead to further applications.\\

\noindent
PACS 04.20.-q, 02.40.-k, 42.15.Eq, 77.84.-s
\end{abstract}

\newpage

\section{Introduction}

Modern metamaterials offer remarkable control over 
electromagnetic fields \cite{PSS}
with applications ranging from future invisibility devices
\cite{PSS}-\cite{Milton}
to existing perfect lenses 
\cite{Pendry}-\cite{Veselago}.
Imagine there were no practical limits on the electromagnetic 
properties of materials.
Given a desired function, how do we find the design
of the device that turns this function into fact?
In this paper, we show that general relativity
provides clear recipes for calculating the
required material properties.
The practical use of 
general relativity in electrical engineering
may seem surprising:
relativity has been associated with 
the physics of gravitation \cite{Telephone}
and cosmology \cite{Peacock}
or, in engineering \cite{VanBladel}, 
has been considered a complication,
not a simplification.
But the design concept described here
is rooted in a simple idea
with a distinguished history:
according to Fermat's Principle 
\cite{LeoNotes,BornWolf,Feynman}
light rays follow the shortest optical paths in media;
they are effective geodesics,
and general relativity has developed the 
theoretical tools for fields in curved geometries
\cite{Telephone}.
Of course, here we exploit only some of the aspects
of general relativity:
we use the physics of curved space, 
but not the physics \cite{Telephone}
that creates space-time curvature
in gravity.\footnote{As
in other analogue models of gravity \cite{Artificial},
we use the kinematic aspects of general relativity,
not the dynamic ones.}
In our case, electromagnetic metamaterials, not masses, 
create effective geometries.
Our theory generalizes the concept
behind the proposed perfect invisibility devices 
\cite{PSS,SPS}
to magneto-electric or moving media
and it incorporates negative refraction 
\cite{Pendry}-\cite{Veselago}.
For example, 
using a simple pictorial argument we show how
to design magnifying perfect lenses.
Finally, metamaterials may be also applied 
for laboratory analogs of general relativity,
in particular for artificial black holes
\cite{Artificial}-\cite{SchUh2}.
In this way, we unify and generalize a range of 
physical phenomena that rely on the geometry of media
and the recent opportunities of metamaterials.
 
Metamaterials are materials with designed properties
that stem from structure, not substance;
where man-made structures determine 
the electromagnetic properties,
structures that are smaller 
than the electromagnetic wavelengths involved.
Metamaterials have a long history:
mediaeval ruby glass, for example, is a metamaterial.
Ruby glass contains nano-scale gold colloids that render
the glass neither golden nor transparent, but ruby,
depending on the size and concentration of the gold droplets.
The color originates from a resonance of the surface plasmons
\cite{Barnes} on the metallic droplets.
Metamaterials {\it per se} are nothing new:
what is new is the degree of control over the structures
in the material that generate the desired properties. 
For example, in the modern version of
negatively-refracting ruby glass \cite{Grigorenko} 
nano-manufactured pairs of gold-pillars 
on a silicon substrate generate finely-tuned plasmon resonances
where each pair acts like a designer atom 
with controllable properties.

Our starting point is not new either:
in the early 1920's
Gordon \cite{Gordon} noticed that moving isotropic media
appear to electromagnetic fields
as certain effective space-time geometries.
Tamm \cite{Tamm1,Tamm2} 
generalized this geometric approach
to anisotropic media
and briefly applied this theory \cite{Tamm2}
to the propagation of light in curved geometries.
In 1960 Plebanski \cite{Plebanski}
formulated the electromagnetic effect 
of curved space-time or curved coordinates 
in concise constitutive equations.
Dielectric media act on electromagnetic fields
as geometries and geometries act as effective media.
In 2000 it was understood \cite{LeoGeometry}
that the dipole forces of electromagnetic fields in media
appear as the effects of geometries as well,
electromagnetic fields act as geometries on media,
a concept used to identify the conditions
for the Abraham or the Minkowski momentum in media 
\cite{LeoMomentum}.
Only very recently 
\cite{PSS}-\cite{LeoNotes}
meta-material implementations of coordinate transformations
were considered as engineering tools,
ideas we take further here:
we show that general relativity
lends the most natural recipes
for such engineering applications.

\section{Theory}

\subsection{Electromagnetism in curved coordinates}

A geometry is characterized by the space-time measure \cite{LL2}
$\mathrm{d}s^2 = 
\sum_{\alpha\beta}g_{\alpha\beta} 
\mathrm{d}x^\alpha \mathrm{d}x^\beta$,
the metric, 
where we denote the space-time coordinates by $x^\alpha$
with Greek indices running from $0$ to $3$.
Latin indices indicate the spatial coordinates and run from $1$ to $3$,
whereas $x^0=ct$ describes time measured in spatial units
with $c$ being the speed of light in vacuum.
The matrix $g_{\alpha\beta}$, the metric tensor,
may vary as a function of the coordinates,
because space-time may be curved or because curved coordinates
are used in flat space.
The determinant $g$ of $g_{\alpha\beta}$ measures
the 4D volume of an infinitesimal space-time element as
$\sqrt{-g}$ times the infinitesimal coordinate volume 
\cite{Telephone,LL2}. 
We denote the matrix inverse of $g_{\alpha\beta}$ by 
$g^{\alpha\beta}$; 
where, as customary in general relativity 
\cite{Telephone,LL2},
the position of the indices indicates that $g_{\alpha\beta}$
is co-variant and $g^{\alpha\beta}$ is contra-variant
under coordinate transformations.
For example, flat space in cylindrical coordinates
$ct$, $r$, $\varphi$, $z$ is described by the metric tensor
$g_{\alpha\beta}=\mathrm{diag}(1,-1,-r^2,-1)$
with matrix inverse 
$g^{\alpha\beta}=\mathrm{diag}(1,-1,-r^{-2},-1)$
and determinant $g=-r^2$.

As our starting point, we use the result of general relativity
\cite{Plebanski,SchleichScully},
derived in Appendix A,
that the free-space Maxwell equations
can be written in the macroscopic form \cite{Jackson}
\begin{equation}
\quad \nabla\cdot\mathbf{D} = \rho \,,\quad 
\nabla\times\mathbf{H} = \frac{\partial \mathbf{D}}{\partial t}
+ \mathbf{j} \,,\quad
\nabla\cdot\mathbf{B} = 0  \,,\quad 
\nabla\times\mathbf{E} = -\frac{\partial \mathbf{B}}{\partial t}
\,,
\label{eq:maxwell}
\end{equation}
or, in Cartesian components,
\begin{eqnarray}
\sum_i\frac{\partial D^i}{\partial x^i} = \rho &,&\quad
\sum_i \frac{\partial B^i}{\partial x^i} = 0 \,,
\label{eq:maxdiv}
\\
\sum_{jk}\epsilon^{ijk}\frac{\partial H_k}{\partial x^j} 
= \frac{\partial D^i}{\partial t} + j^i
&,&\quad
\sum_{jk}\epsilon^{ijk}\frac{\partial E_k}{\partial x^j} 
= -\frac{\partial B^i}{\partial t}
\,,
\label{eq:maxcurl}
\end{eqnarray}
where $\epsilon^{ijk}$ is the completely antisymmetric 
Levi-Civita tensor \cite{Telephone}: 
in Cartesian components $\epsilon^{ijk}$ is $1$ 
for all cyclic permutations 
of $\epsilon^{123}$, $-1$ for all cyclic permutations of
$\epsilon^{213}$ and $0$ otherwise.
The spatial indices indicate that in this representation
$E_i$ and $H_i$ form the components of vectors
that are co-variant under purely spatial transformations,
whereas $D^i$ and $B^i$ constitute contra-variant
spatial vectors.
The charge density $\rho$ and the current density $\mathbf{j}$
are given by 
$\sqrt{-g}\,j^0$ and $c\sqrt{-g}\,j^i$
of the four-current $j^\alpha$ \cite{LL2}.
In empty but possibly curved space,
the electromagnetic fields are connected by the 
constitutive equations in SI units \cite{Plebanski},
\begin{equation}
\mathbf{D} = \varepsilon_0\varepsilon\,\mathbf{E}
+ \frac{\mathbf{w}}{c}\times\mathbf{H}
\,,\quad
\mathbf{B} = \frac{\mu}{\varepsilon_0 c^2}\,\mathbf{H}
- \frac{\mathbf{w}}{c}\times\mathbf{E}
\,.
\label{eq:constitution}
\end{equation}
In dielectric media, the $\mathbf{E}$, $\mathbf{B}$ vectors
generate electric polarizations and magnetizations
that constitute the $\mathbf{D}$, $\mathbf{H}$ fields.
The constitutive equations are described here in implicit form, 
but
one can also express them as $\mathbf{D}$, $\mathbf{H}$
as functions of $\mathbf{E}$ and $\mathbf{B}$ 
\cite{VanBladel}.
In general, the electric permittivity $\varepsilon$ 
and the magnetic permeability $\mu$ 
are symmetric matrices
-- space-time appears as an anisotropic medium --
but, in empty space, $\varepsilon$ and $\mu$
are equal, as in perfect impedance
matching \cite{Jackson}.
The $\mathbf{w}$ vector describes a
magneto-electric coupling between the magnetic
and the electric fields.
Some materials are magneto-electric, 
but the simplest example is a moving medium
\cite{VanBladel,LeoGeometry,LL8}, because
a moving medium responds to the electromagnetic field
in locally co-moving frames and
Lorentz transformations from the laboratory frame
naturally mix electric and magnetic fields
\cite{VanBladel,LL8}.
In explicit form \cite{Plebanski}, 
\begin{equation}
\varepsilon = \mu = -\frac{\sqrt{-g}}{g_{00}}\,g^{ij}
\,,\quad 
w_i = \frac{g_{0i}}{g_{00}}
\label{eq:constrel}
\,,
\end{equation}
the $\varepsilon$ and $\mu$ are 
constructed from the spatial components
of the inverse metric tensor and from the determinant and the
time-time component of the metric tensor,
whereas the $\mathbf{w}$ vector is given in terms 
of the time-space components of the metric tensor.
Empty space appears as an impedance-matched
anisotropic magneto-electric or moving medium.

\subsection{Transformation media}

\begin{figure}[h]
\begin{center}
\includegraphics[width=24.0pc]{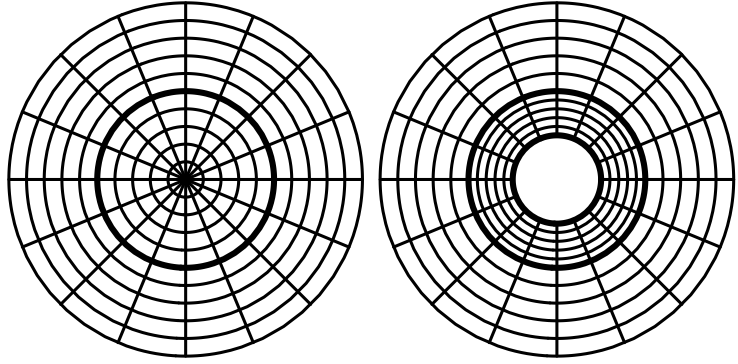}
\caption{
\small{
Transformation media implement coordinate transformations.
The left figure shows an orthogonal grid of coordinates
in electromagnetic space 
(a slice of cylindrical coordinates at constant $z$).
The right figure shows the transformed grid in physical
space that corresponds to the invisibility device
described by Eq.\ (\ref{eq:invisibility}) and
illustrated in Fig.\ \ref{fig:invisibility}. 
}
\label{fig:coordinates}}
\end{center}
\end{figure}

Since empty space appears as a medium,
what happens if a medium appears as empty space?
This is the idea behind the recent proposals
for invisibility devices 
\cite{PSS}-\cite{LeoNotes}.
To be more precise,
suppose that the medium appears as the result
of a coordinate transformation from some fictitious space-time,
say electromagnetic space-time, to physical space,
see, for example, Fig.\ \ref{fig:coordinates}.
Electromagnetic space-time could be flat with light traveling
along straight lines, 
whereas to electromagnetic fields physical space appears to be
curved, bending light.
Of course, this apparent curvature is an illusion,
the same type of illusion as straight lines 
appearing curved in curved coordinates,
because in theory it is removable by the 
inverse coordinate transformation:
but, in practice, one can exploit this apparent curvature
to create illusions 
\cite{PSS}-\cite{LeoNotes}.

We use primes to denote the geometry and coordinates 
of electromagnetic space-time\footnote{We 
use the opposite notation of Ref.\ \cite{PSS},
since, we believe, it is more logical.}
and describe physical space in generalized 
coordinates $x^i$ 
with spatial metric $\gamma_{ij}$ and determinant $\gamma$,
for keeping the theory as flexible as possible.  
For example, we may wish to use cylindrical coordinates
$r$, $\varphi$, $z$ with spatial metric tensor
$\gamma_{ij}=\mathrm{diag}(1,r^2,1)$ and $\gamma=r^2$.
The metric $\gamma_{ij}$ should
differ from the effective geometry $g_{\alpha\beta}$
generated by the medium.
The transformation rules of tensors in general relativity
give rise to a simple recipe for the construction of media
that facilitate space-time coordinate transformations.
Contra-variant tensors are transformed as  
\cite{VanBladel,Telephone, LL2}
\begin{equation}
g^{\alpha\beta} = \sum_{\alpha'\beta'}
\frac{\partial x^\alpha}{\partial x'^{\alpha'}}
\frac{\partial x^\beta}{\partial x'^{\beta'}}\,
g'^{\alpha'\beta'}
\label{eq:rule}
\,.
\end{equation}
The transformed inverse metric serves as the building block
of the dielectric tensors (\ref{eq:constrel}) of the medium.
If we wish to express physical space in generalized coordinates, 
we need to consider a subtlety
that appears when we write the divergences
in Maxwell's equations (\ref{eq:maxdiv})
in spatially co-variant form \cite{LL2}
\begin{equation}
\frac{1}{\sqrt{\gamma}}
\sum_i\frac{\partial \sqrt{\gamma}D^i}{\partial x^i} = \rho 
\,,\quad
\frac{1}{\sqrt{\gamma}}
\sum_i\frac{\partial \sqrt{\gamma}B^i}{\partial x^i} =  0
\,.
\end{equation}
The $\varepsilon^{ij}$ and $\mu^{ij}$ tensors
are naturally contra-variant with respect to the background 
geometry of physical space, but $\gamma$
differs from $-g$, in general.
Consequently, 
for writing the medium as an active coordinate transformation,
we need to multiply $\mathbf{D}$ and $\mathbf{B}$
by $\sqrt{\gamma}$ in the constitutive equations 
(\ref{eq:constitution})
and re-scale $\rho$ and $\mathbf{j}$ accordingly, 
which is also consistent with the form
of the Levi-Civita tensor in curved coordinates \cite{Telephone}
in the curls in Maxwell's equations (\ref{eq:maxcurl}). 
However, a second subtlety arises from the curls:
the coordinate transformation may turn 
a right-handed coordinate system into a locally left-handed one,
but the curls (\ref{eq:maxcurl})
implicitly assume a right-handed system,
because $\epsilon^{ijk}$
changes sign under coordinate transformations
from right to left-handed systems.
Consequently, we need to invert the sign of $\varepsilon$, $\mu$
and $\rho$, $\mathbf{j}$ wherever this is the case. 
The transformation to a left-handed coordinate system
thus corresponds to negative refraction \cite{Smith,Veselago}
occurring in what has, for other reasons  \cite{Veselago},
been fittingly described as left-handed media.\footnote{Negative
refraction in left-handed media 
is not related to the case of negative
phase velocities in gravity \cite{Lakhtakia,McCall},
because transformations to left-handed or any other
coordinates have no physical significance there,
in contrast to transformations facilitated by media.}
Taking all this into account, we arrive at the simple recipe
\begin{equation}
\varepsilon^{ij} = \mu^{ij} = 
\mp\frac{\sqrt{-g}}{\sqrt{\gamma}}\,\frac{g^{ij}}{g_{00}}
\,,\quad 
w_i = \frac{1}{\sqrt{\gamma}}\, \frac{g_{0i}}{g_{00}}
\label{eq:recipe}
\,.
\end{equation}
The $\mp$ sign indicates the handedness:
minus for right-handed transformations and
plus for locally left-handed ones.
Starting from the inverse metric $g'^{\alpha\beta}$
in electromagnetic space,
the $g^{\alpha\beta}$ matrix is calculated 
according to the transformation rule (\ref{eq:rule}).
The matrix inverse of $g^{\alpha\beta}$ 
gives the metric tensor $g_{\alpha\beta}$
and the inverse of the determinant of $g^{\alpha\beta}$
gives $g$.
Equations (\ref{eq:recipe}) specify the required
electromagnetic properties that will turn 
the coordinate transformation into reality.
In the special case of right-handed and purely spatial
transformations our recipe agrees with the theory
of Refs.\ \cite{PSS,Ward}, see Appendix B,
where however, instead of the contra-variant 
$\varepsilon$ and $\mu$ mixed tensors occur
with one of the indices lowered by the spatial metric,
\begin{equation}
\varepsilon^i_k = \mu^i_k = 
\sum_j \varepsilon^{ij}\gamma_{jk}
\label{eq:mix}
\,.
\end{equation}
Here we consider more general transformations 
that may mix space and time and that may be multivalued.
As long as the matrix (\ref{eq:rule}) is single-valued
and not explicitly time-dependent
such a device is physically allowed and stationary. 

The coordinate transformation encodes the function of the device.
Outside of it, the electromagnetic coordinates agree with the 
coordinates of physical space:
the transformation is trivial; 
$\varepsilon$ and $\mu$
are unity and $\mathbf{w}$ vanishes.
Inside the device,
electromagnetic fields are controlled as prescribed by the 
coordinate transformation. 
If the physical coordinates enclose a hole
one obtains the blueprint of an invisibility device
\cite{PSS}-\cite{LeoNotes}. 
We show in the next section that
perfect lensing 
\cite{Pendry}-\cite{Veselago}, 
the optical Aharonov-Bohm effect \cite{LPStor,LPLiten,Hannay,CFM}
and optical black holes \cite{Artificial}-\cite{SchUh2}
can be understood as applications of the same idea  as well.

\section{Examples}

\subsection{Perfect invisibility devices}

Perfect invisibility devices \cite{PSS,SPS} 
facilitate coordinate transformations with holes in physical space.
In this way, electromagnetic radiation is naturally guided 
around such excluded regions.
Anything placed inside is hidden.
Perfect invisibility devices \cite{PSS,SPS} must employ 
anisotropic media,
because the inverse scattering problem for waves
in isotropic media has unique solutions 
\cite{Nachman,Gbur}.
They also require media where the phase velocity of light
approaches infinity at the inside of the cloaking layer,
because coordinate transformations with holes 
rip apart points of zero volume in electromagnetic space
and tear them to finite volumes in physical space,
unless the coordinate transformation is singular.
To prove this, consider the determinant of the
$\varepsilon$ and $\mu$ tensor (\ref{eq:recipe}).
For an invisibility device \cite{PSS,SPS} only space 
is transformed, so $\mathbf{w}$ vanishes, 
$g_{00}$ is unity and $-g$ is reduced 
to the inverse of the determinant of the spatial components $g^{ij}$.
We obtain from Eqs.\ (\ref{eq:rule}) and (\ref{eq:recipe}) 
in three-dimensional space 
\begin{equation}
\mathrm{det}\,\varepsilon = 
(-g)^{1/2}\gamma^{-3/2} =
(-g')^{1/2}\gamma^{-3/2}\,J \,,\quad
J = \frac{\partial(x'^1,x'^2,x'^3)}{\partial(x^1,x^2,x^3)}
\,,
\end{equation}
in terms of the Jacobian $J$ \cite{LL2}.
At a point of measure zero $\sqrt{-g'}$ vanishes and so does
$\mathrm{det}\,\varepsilon$, the product of the 
eigenvalues of $\varepsilon$ and $\mu$.
Consequently, at least one of the eigenvalues of 
$\varepsilon$ and $\mu$ are zero;
therefore in at least one direction of the anisotropic medium
the phase-velocity of light 
diverges near the inside of the cloak.
The speed of light in media can reach high values
in narrow frequency ranges or, naturally,
for static fields.
On the other hand, invisibility devices that are only perfect
within the limits of geometrical optics 
\cite{LeoConform,LeoNotes,Hendi}
are not subject to such constraints \cite{Hendi}.

\begin{figure}[h]
\begin{center}
\includegraphics[width=20.0pc]{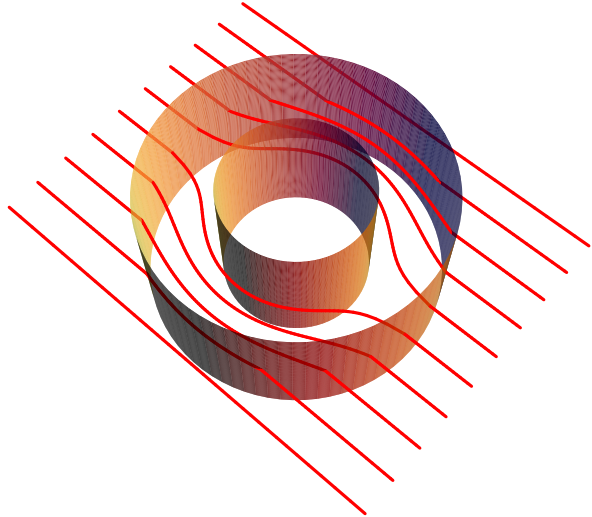}
\caption{
\small{
Invisibility device. 
The transformation medium (\ref{eq:invisibility})
acts as an invisibility cloak,
guiding light around the interior of the cloak
without causing any distortion.
The device facilitates the coordinate transformation
illustrated in Fig.\ \ref{fig:coordinates}.
}
\label{fig:invisibility}}
\end{center}
\end{figure}

Figure \ref{fig:invisibility} illustrates a cylindrical 
invisibility device 
that stretches the $z$ axis out in the radial direction
to a full cylinder of radius $R_1$, 
compressed within a cylindrical volume of radius $R_2$,
as shown in Fig.\ \ref{fig:coordinates}. 
This device is an invisibility cloak of thickness $R_2-R_1$ where 
anything placed inside the inner radius $R_1$ is hidden.
We obtain from the theory 
\begin{equation}
r=R_1 + r'\frac{R_2-R_1}{R_2}
\quad\Longrightarrow\quad
\varepsilon_j^i = \frac{R_2}{R_2-R_1}\,\frac{r'}{r}\,
\mathrm{diag}\left(\frac{(R_2-R_1)^2}{R_2^2},
\frac{r^2}{r'^2},1\right)
\label{eq:invisibility}
\end{equation}
within $R_1\le r \le R_2$ or, equivalently, $0 \le r' \le R_2$.
Clearly, close to the lining of the cloak at $r\rightarrow R_1$
where $r'\rightarrow 0$
the speed of light in the $r$ and $z$ directions diverges,
whereas in $\varphi$ direction the phase velocity
tends to zero.

\subsection{Perfect lenses}

\begin{figure}[h]
\begin{center}
\includegraphics[width=16.0pc]{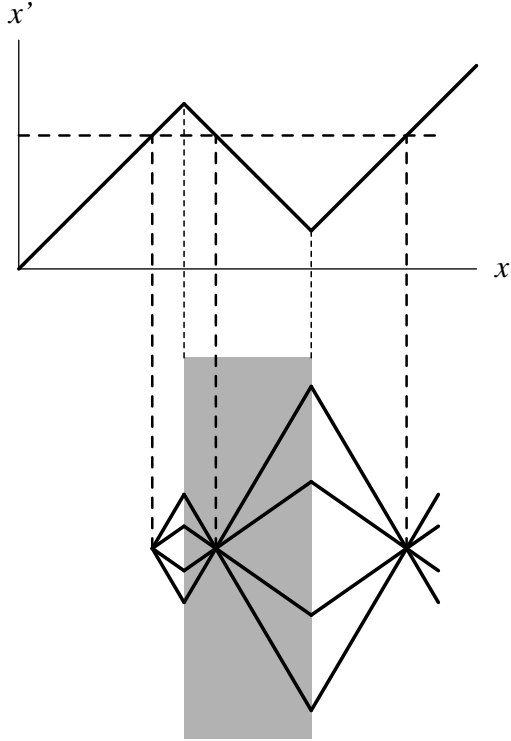}
\caption{
\small{
Perfect lens.
Negatively refracting perfect lenses employ transformation media.
The top figure shows a suitable coordinate transformation
from the physical $x$ axis to the electromagnetic $x'$,
the lower figure illustrates the corresponding device.
The inverse transformation from $x'$ to $x$ is either triple
or single-valued.
The triple-valued segment on the physical $x$ axis 
corresponds to the focal region of the lens: any source point
has two images, one inside the lens and one on the other side.
Since the device facilitates an exact coordinate transformation,
the images are perfect with a resolution below the
normal diffraction limit: the lens is perfect \cite{Pendry}.
In the device,
the transformation changes 
right-handed into left-handed coordinates.
Consequently, the medium employed here is left-handed, 
with negative refraction \cite{Veselago}.
}
\label{fig:lens}}
\end{center}
\end{figure}

In perfect invisibility devices, 
electromagnetic space does not
cover the entire physical space:
here we show that perfect lenses correspond to
multi-valued electromagnetic space.
Consider in Cartesian coordinates
the multi-valued transformation $x(x')$ illustrated 
in Fig.\ \ref{fig:lens}, 
whereas all other coordinates are not changed.
In the fold of the function $x(x')$
a point $x'$ in electromagnetic space 
has three faithful images in physical space.
Obviously, electromagnetic fields 
at one of those points are perfectly imaged onto
the other: the device is a perfect lens \cite{Pendry}.
This simple pictorial argument may contribute to settling
the debate on perfect lensing 
\cite{Minkel}-\cite{P2Re}
in addition to the experimental proof for 
enhanced imaging in existing perfect lenses
\cite{Grbic}-\cite{Melville}. 
Our theory also shows why perfect lensing
requires left-handed media with negative 
$\varepsilon$ and $\mu$ \cite{Smith,Veselago}:
inside the device, {\it i.e.} inside of the $x'$ fold,
the derivative of $x(x')$ becomes negative and
the coordinate system changes handedness.
We obtain from our recipe (\ref{eq:recipe})
the compact result
\begin{equation}
\varepsilon = \mu = 
\mathrm{diag}
\left(\frac{\mathrm{d}x'}{\mathrm{d}x},
\frac{\mathrm{d}x}{\mathrm{d}x'},
\frac{\mathrm{d}x}{\mathrm{d}x'}\right)
\,.
\end{equation}
If, for example, $\mathrm{d}x'/\mathrm{d}x$
is $-1$ inside the device and $+1$ outside,
we obtain the standard perfect lens 
\cite{Pendry}-\cite{Veselago} based
on an isotropic left-handed material with  
$\varepsilon=\mu=-1$ inside;
but this is not the most general choice.
One could use an anisotropic medium to magnify
perfect images, in contrast to the existing
examples \cite{Grbic}-\cite{Melville}, by embedding the
source or the image in transformation media with 
$|\mathrm{d}x'/\mathrm{d}x| \neq 1$.

\subsection{Optical Aharonov-Bohm effect}

\begin{figure}[t]
\begin{center}
\includegraphics[width=30.0pc]{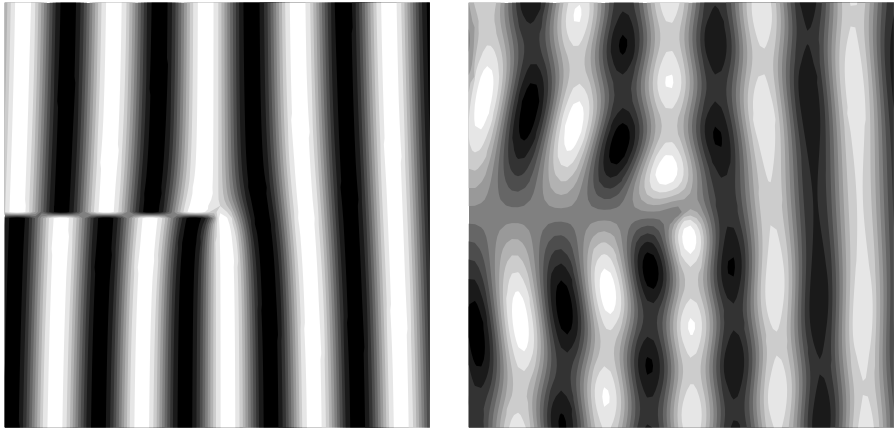}
\caption{
\small{
Aharonov-Bohm effect.
A fluid vortex generates the optical Aharonov-Bohm effect
described by the coordinate transformation (\ref{eq:ab}).
Light, incident from the right, 
is Fresnel-dragged by the moving medium:
light propagating with the flow is advanced,
whereas light propagating against the current is retarded.
The wave should develop the phase slip shown in the left figure.
However, although the transformation (\ref{eq:ab}) is exact,
physical space-time is multi-valued here.
Instead of the simple phase slip,
the light turns out to exhibit the characteristic interference
pattern illustrated in the right figure
\cite{AB,Peshkin}.
}
\label{fig:ab}}
\end{center}
\end{figure}

Perfect invisibility devices and perfect lenses exploit
transformation media with non-trivial topology,
with excluded regions in physical space  
or folds in electromagnetic space.
Here we consider an example where physical space
is multi-valued, but the medium is single-valued 
and hence physically allowed:
the optical Aharonov-Bohm effect \cite{LPStor,LPLiten,Hannay,CFM}.
One can demonstrate this effect with light
passing through a water vortex \cite{LPStor}
or with slow light in Bose-Einstein condensates \cite{LPLiten}.
Note that the effect is related to 
the Aharonov-Bohm effect with surface waves
\cite{Berryetal}-\cite{Vivanco}
and to the gravitational
Aharonov-Bohm effect \cite{Stachel}.
The optical Aharonov-Bohm effect is an example
of a transformation medium that mixes space and time, 
and yet the medium is stationary.
Consider in cylindrical coordinates the transformation
\begin{equation}
ct = ct' + a\varphi' \,,\quad
r = \frac{r'}{n} \,,\quad
\varphi = \varphi' \,,\quad
z = \frac{z'}{n}
\label{eq:ab}
\end{equation}
with the constants $n$ and $a$.
We obtain from our theory (\ref{eq:recipe}) that
the medium is isotropic with refractive index $n$
and is magneto-electric with $w_i=(0,a/r,0)$,
which corresponds
to a fluid vortex with velocity profile 
$\mathbf{u}/c=\mathbf{w}/(n^2-1)$,
as we see comparing Eq.\ (\ref{eq:constitution})
with the constitutive equations
of moving media \cite{VanBladel,LL8} 
in lowest relativistic order.
Normally, 
a moving medium Fresnel-drags light \cite{LPStor},
but in the case of a vortex,
light rays follow straight lines,
because the transformation (\ref{eq:ab})
changes only time, such that straight rays
in electromagnetic space are mapped onto straight lines
in physical space.
Similarly, in the original 
Aharonov-Bohm effect \cite{AB,Peshkin}
electron rays are not bent by a magnetic vortex,
but they develop a phase slip in the direction of incidence.
In our case, when the light has passed the vortex,
the time change in the transformation (\ref{eq:ab})
results in a phase slip of $\pm \pi ak$, 
where $k$ is the wave number,
depending on whether the light propagates with
or against the current,
see Fig.\ \ref{fig:ab}.
Physical space-time is multi-valued,
with a branch cut in the direction of incidence,
resembling the infinitely sheeted Riemann 
surfaces around a logarithmic branch point \cite{Ablowitz},
but the medium is single-valued 
and has the simple physical interpretation of 
a moving fluid forming a vortex.

\subsection{Artificial black holes}

\begin{figure}[t]
\begin{center}
\includegraphics[width=25.0pc]{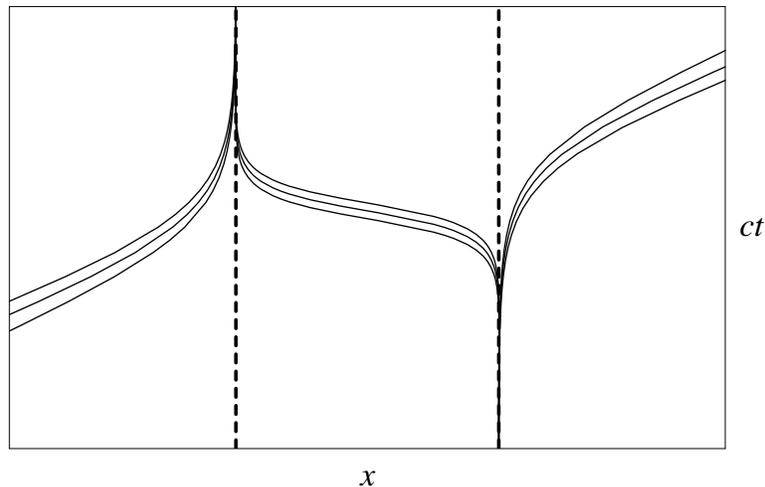}
\caption{
\small{
Event horizons.
Space-time diagram showing light propagation
near two artificial event horizons,
a white-hole horizon at the left and a black-hole
horizon at the right.
The horizons could be made by a
medium moving from the right side to the left.
The places $x$ where the medium velocity 
reaches the local speed of light in the medium are the horizons.
At the black-hole horizon the medium turns from 
subluminal to superluminal velocity;
at the white-hole horizon it returns to subluminal speed.
The exponential confinement near the horizons
translates into exponential red shifts 
of optical oscillations at the black hole
and exponential blue shifts at the white hole.
Magneto-electric media could mimic the 
moving medium generating the horizons.
}
\label{fig:horizon}}
\end{center}
\end{figure}

Moving isotropic media generate the effective space-time
geometry discovered by Gordon in 1923 \cite{Gordon}.
Consider an effectively one-dimensional situation
where the medium is moving in $x$ direction
and the electromagnetic field varies along the $x$ axis
with field vectors pointing orthogonal to $x$.
An impedance-matched medium of refractive
index $n$ and velocity $u$ is described by the
inverse metric tensor \cite{Gordon,LeoGeometry,LPStor}
\begin{equation}
g^{\alpha\beta} =
\left(
\begin{array}{cccc}
1 & 0 & 0 & 0\\
0 & -1& 0 & 0\\
0 & 0 & -1 & 0\\
0 & 0 & 0 & -1
\end{array}
\right)
+ \frac{n^2-1}{1-u^2/c^2}
\left(
\begin{array}{cccc}
1 & u/c & 0 & 0\\
u/c & u^2/c^2 & 0 & 0\\
0 & 0 & 0 & 0 \\
0 & 0 & 0 & 0 \\
\end{array}
\right)
\,.
\label{eq:gordon}
\end{equation}
Both the velocity $u$ and the index $n$ may vary.
We show that this effective geometry is generated 
from empty space
by the coordinate transformation
\begin{equation}
ct' = \frac{c}{2}(t_- + t_+) \,,\quad
x' = \frac{c}{2}(t_- - t_+) \,,\quad
t_\pm = t - \int \frac{n \pm u/c}{nu \pm c}\,\mathrm{d}x
\,,
\label{eq:moving}
\end{equation}
as long as we restrict our attention to fields varying in $x$ 
direction.
For this, we use the transformation rule (\ref{eq:rule}),
but from un-primed to primed space-time,
and find that 
\begin{equation}
g'^{\alpha\beta} =
\mathrm{diag}\left(\frac{n^2(c^2-u^2)}{c^2-n^2u^2},
-\frac{n^2(c^2-u^2)}{c^2-n^2u^2},-1,-1\right)
\,.
\end{equation}
In electromagnetic space-time,
we are free to apply the recipe (\ref{eq:recipe})
for assigning the medium properties as well as in physical space.
We find
$\varepsilon'=\mu'=1$ 
in the directions orthogonal to $x$
and $w'=0$: vacuum,
which proves our point.
Here wave packets are superpositions 
of waves propagating in positive or
negative direction as functions of either $t_+$ or $t_-$.
Substituting for $t_\pm$ 
the representations (\ref{eq:moving})
as functions of $ct$ and $x$,
we obtain the electromagnetic waves in physical space.
The transformation (\ref{eq:moving})
describes the relativistic addition theorem of velocities
\cite{VanBladel,LL2}
for the medium and light 
propagating in positive or negative direction.
Interesting phenomena occur when the velocity 
of the medium, $u$, 
reaches the speed of light in the medium, $c/n$,
which is possible in theory and perhaps also in practice.
In this case, 
the integral in $t_\pm$ develops a logarithmic singularity.
Light propagating against the current freezes with
exponentially increasing oscillations
at a horizon \cite{Brout}, see Fig.\ \ref{fig:horizon}.
This horizon is completely analogous to the event horizon
of the black hole  
if the light is escaping from a superluminal region 
and to a white hole if the light is attempting to
enter a counter-propagating superluminal medium
\cite{LeoReview}.
Horizons cut physical space-time into distinct regions
without possible communication. 
They correspond to disconnected branches of multi-valued
electromagnetic space, covering it multiple times,
in contrast to the case of perfect lensing
where the folds are connected.
Horizons are predicted to exhibit remarkable quantum effects
\cite{Brout}, in particular Hawking radiation \cite{Hawking},
that are extremely difficult to observe in astronomy,
but may possibly be demonstrated in 
laboratory analogs
\cite{Artificial}-\cite{SchUh2}.
Our method suggests a magneto-electric
analog of the event horizon, perhaps with metamaterials,
if we interpret the effective geometry (\ref{eq:gordon})
as generating the constitutive equations
(\ref{eq:constitution}) 
expressed in terms of the $D$ and $H$ fields as
\begin{equation}
D_y = \varepsilon_0 \frac{(n^2-u^2/c^2)E_y-(n^2-1)uB_z}
{n(1-u^2/c^2)}
\,,\quad
H_y = \varepsilon_0 \frac{(c^2-n^2u^2)B_y-(n^2-1)uE_z}
{n(1-u^2/c^2)}
\,. \label{bhcons}
\end{equation}
Here the functions $n$ and $u$ of the moving medium
appear as parameters of a magneto-electric material at rest.
The material establishes horizons
at $u=c/n$ without singular dielectric properties,
in contrast to the previous proposal \cite{Reznik}.
general relativity can be put to practical use 
in electrical engineering,
but electrical engineering may be also used for demonstrating
some elusive effects of general relativity.

\section{Conclusions}

Perfect invisibility devices \cite{PSS,SPS}, 
perfect lenses 
\cite{Pendry}-\cite{Veselago},
the optical Aharonov-Bohm effect 
\cite{LPStor,LPLiten,Hannay,CFM}
and artificial event horizons 
\cite{Artificial}-\cite{SchUh2}
are all examples of one unifying concept:
electromagnetic media 
that facilitate coordinate transformations.
Adopting ideas from general relativity 
\cite{Gordon}-\cite{SchleichScully}
we developed a concise formalism for finding
the  properties of meta-materials
that turn a desired function into fact.
We extended the previously reported method 
\cite{PSS,SPS}
to media that act as space-time transformations
and that may exhibit negative refraction
and signs of multi-valued physical space.
The most interesting properties 
of such transformation media 
seem to stem from non-trivial topologies:
spaces with holes in the case of invisibility devices,
coordinate folds for negative refraction,
multi-valued physical space for the Aharonov-Bohm effect
and multiple sheets of electromagnetic space-time
in the case of artificial event horizons.
Our theory can be extended in at least two directions.
So far we assumed that electromagnetic
space-time is empty, but one could easily incorporate
media with non-trivial $\varepsilon'$ and $\mu'$ here.
We also assumed that the mapping between 
electromagnetic fields in real and fictitious space is exact. 
In practice, the accuracy of geometrical optics \cite{BornWolf}
is often completely sufficient;
one could further generalize our method to 
include transformations that are only exact within the
validity range of geometrical optics.
Optical conformal mapping \cite{LeoConform,LeoNotes}
is such an example.
These more general transformation media 
act as local transformations of the dispersion relation
of light waves.
They all share the same spirit:
electromagnetic media act as effective geometries.

\section*{Acknowledgments}
We thank
John Allen,
Ildar Gabitov,
Andrew Green,
Chris Hooley,
Natalia Korolkova
and
Irina Leonhardt
for their help and helpful comments.
Our work was supported by the Leverhulme Trust and 
the Engineering and Physical Sciences Research Council.

\newpage

\renewcommand{\theequation}{A\arabic{equation}}
\setcounter{equation}{0}

\section*{Appendix A}
In this appendix we show that the free-space Maxwell equations 
in generally covariant form are equivalent 
to Maxwell's equations in a material medium 
with constitutive equations 
(\ref{eq:constitution}),(\ref{eq:constrel}). 
The generally covariant Maxwell equations are 
\cite{Telephone,LL2}
\begin{equation}
F_{[\mu\nu;\lambda]}=\partial_{[\lambda}F_{\mu\nu]}=0
\,,\quad
\varepsilon_0F^{\mu\nu}_{\ \ \ ;\nu}=
\frac{\varepsilon_0}{\sqrt{-g}}\partial_{\nu}
\left(\sqrt{-g}F^{\mu\nu}\right)=j^\mu \,, 
\label{amax} 
\end{equation}
where $F_{\mu\nu}$ denotes the electromagnetic field tensor, 
$j^\mu$ is the four-current, 
the square brackets denote antisymmetrization and 
the semi-colon indicates covariant differentiation. 
We employ Einstein's summation convention over repeated indices.
The covariant tensor $F_{\mu\nu}$ contains 
the $\mathbf{E}$ and $\mathbf{B}$ fields in the usual manner:
\begin{equation} \label{aF}
 F_{\mu\nu}=  \begin{pmatrix}
       0 &\  \ E_x & \ \ E_y & \ \ E_z \\
      -E_x & \ \ 0 & \ \ -cB_z & \ \ cB_y \\
      -E_y & \ \ cB_z & \ \ 0 & \ \ -cB_x \\
      -E_z & \ \ -cB_y & \ \ cB_x & \ \ 0  
   \end{pmatrix}.
\end{equation}
We define a quantity $H^{\mu\nu}$ with contravariant indices by
\begin{gather} 
H^{\mu\nu}=\varepsilon_0\sqrt{-g}F^{\mu\nu}=
\varepsilon_0\sqrt{-g}g^{\mu\lambda}g^{\nu\rho}F_{\lambda\rho}  
\label{acons1}  \\[8pt]
\Longrightarrow\quad F_{\mu\nu}=
\frac{1}{\varepsilon_0
\sqrt{-g}}g_{\mu\lambda}g_{\nu\rho}H^{\lambda\rho}  
\label{acons2}
\end{gather}
and regard $H^{\mu\nu}$ 
as being constructed 
from $\mathbf{D}$ and $\mathbf{H}$ fields as follows:
\begin{equation} \label{aH}
 H^{\mu\nu}=  \begin{pmatrix}
       0 &\  \ -D_x & \ \ -D_y & \ \ -D_z \\
      D_x & \ \ 0 & \ \ -H_z/c & \ \ H_y/c \\
      D_y & \ \ H_z/c & \ \ 0 & \ \ -H_x/c \\
      D_z & \ \ -H_y/c & \ \ H_x/c & \ \ 0  
   \end{pmatrix}.
\end{equation}
Then, introducing a new four-current $J^\mu=\sqrt{-g}j^\mu$, 
Maxwell's equations (\ref{amax}) can be written
\begin{equation}
\partial_{[\lambda}F_{\mu\nu]}=0, 
\qquad  \partial_{\nu}H^{\mu\nu}=J^\mu \,,
\end{equation}
which are the electromagnetic equations in a material medium,
described by the constitutive equations (\ref{aF})-(\ref{aH}),
with free charge and current densities $J^\mu$.

To obtain relations between the vector fields 
$\mathbf{D}$, $\mathbf{H}$ and 
$\mathbf{E}$, $\mathbf{B}$ 
consider first the components $F_{0i}$; 
from Eqs.\ (\ref{aF}), (\ref{acons2}) and (\ref{aH}) 
one obtains
\begin{equation} \label{aecon}
E_i=\frac{1}{\varepsilon_0\sqrt{-g}}
\left(g_{i0}g_{j0}-g_{ij}g_{00}\right) D_j-
\frac{c}{\varepsilon_0\sqrt{-g}}[jkl]g_{0j}g_{ik}\,H_l\,,
\end{equation}
where $[jkl]$ denotes the completely antisymmetric permutation 
symbol \cite{Telephone} with $[xyz]=+1$. 
We simplify this result as follows.
The identity
\[
g_{\mu\lambda}g^{\lambda\nu}=\delta_{\mu}^{\nu}
\]
gives
\begin{gather}
g_{i\lambda}g^{\lambda 0}=0 
\quad \Longrightarrow \quad  
g_{i0}=-\frac{1}{g^{00}}\,g_{ij}g^{j0} \,, 
\label{agid1} \\[8pt]
g_{0\lambda}g^{\lambda i}=0 
\quad \Longrightarrow \quad  
g^{i0}=-\frac{1}{g_{00}}\,g^{ij}g_{j0} \,, 
\label{agid2} \\[8pt]
g_{j\lambda}g^{\lambda i}=
g_{j0}g^{0 i}+g_{jk}g^{ki}=\delta^i_j \,.  
\label{agid3}
\end{gather}
Use of Eqs.\ (\ref{agid1}) or (\ref{agid2}) 
in Eq.\ (\ref{agid3}) produces the two relations
\begin{equation}
\left(g^{ij}-\frac{1}{g^{00}}g^{i0}g^{k0}\right)g_{kj}=
\delta^i_j \,,\quad
g^{ik}\left(g_{kj}-\frac{1}{g_{00}}g_{k0}g_{j0}\right)=
\delta^i_j \,, \label{agid5}
\end{equation}
which reveal inverse-related $3\times 3$ matrices. 
In view of Eqs.\ (\ref{agid3}) and (\ref{agid5}),  
transvection of (\ref{aecon}) by $g^{li}$ yields
\begin{equation} \label{aDcons}
D_i=-\frac{\varepsilon_0
\sqrt{-g}}{g_{00}}g^{ij}E_j+
\frac{\varepsilon_0c}{g_{00}}[ijk]g_{j0}H_k \,,
\end{equation}
the first of the constitutive equations 
(\ref{eq:constitution}) with (\ref{eq:constrel}).

To obtain the second constitutive relation, 
we employ the tensors dual to $F_{\mu\nu}$ and $H^{\mu\nu}$. 
This requires use of the 4D Levi-Civita tensor \cite{Telephone}, 
given by
\begin{equation}
\epsilon_{\mu\nu\lambda\rho}=\sqrt{-g}\,[\mu\nu\lambda\rho]
\,,\quad 
\epsilon^{\mu\nu\lambda\rho}=
-\frac{1}{\sqrt{-g}}\,[\mu\nu\lambda\rho]
\,,\quad 
[0123]=+1 \,.
\end{equation}
The dual tensors $^*\!F^{\mu\nu}$ and $^*\!H_{\mu\nu}$ 
are defined by \cite{Telephone}
\begin{gather}
^*\!F^{\mu\nu}=
\frac{1}{2}\epsilon^{\mu\nu\lambda\rho}F_{\lambda\rho} 
\qquad \Longrightarrow \qquad
F_{\mu\nu}=
\frac{1}{2}\epsilon_{\mu\nu\lambda\rho}\,^*\!F^{\lambda\rho}, 
\\[8pt]
^*\!H_{\mu\nu}=
\frac{1}{2}\epsilon_{\mu\nu\lambda\rho}H^{\lambda\rho} 
\qquad \Longrightarrow \qquad
H^{\mu\nu}=
\frac{1}{2}\epsilon_{\mu\nu\lambda\rho}\,^*\!H^{\lambda\rho},
\end{gather}
so they have components
\begin{gather} 
 ^*\!F^{\mu\nu}=  \frac{1}{\sqrt{-g}}\begin{pmatrix}
       0 &\  \ -cB_x & \ \ -cB_y & \ \ -cB_z \\
      cB_x & \ \ 0 & \ \ E_z & \ \ -E_y \\
      cB_y & \ \ -E_z & \ \ 0 & \ \ E_x \\
      cB_z & \ \ E_y & \ \ -E_x & \ \ 0  
   \end{pmatrix}, \label{adual1} \\[8pt]
 ^*\!H_{\mu\nu}=  \sqrt{-g}\begin{pmatrix}
       0 &\  \ H_x/c & \ \ H_y/c & \ \ H_z/c \\
      -H_x/c & \ \ 0 & \ \ D_z & \ \ -D_y \\
      -H_y/c & \ \ -D_z & \ \ 0 & \ \ D_x \\
      -H_z/c & \ \ D_y & \ \ -D_x & \ \ 0  
   \end{pmatrix}. \label{adual2}
\end{gather}
Re-expressed in terms of the dual tensors, 
the constitutive equations (\ref{acons1})-(\ref{acons2}) read
\begin{gather}
^*\!H_{\mu\nu}=
\varepsilon_0\sqrt{-g}g_{\mu\lambda}g_{\nu\rho}
\,^*\!F^{\lambda\rho},  \label{acons3}  \\[8pt]
^*\!F^{\mu\nu}=
\frac{1}{\varepsilon_0\sqrt{-g}}
g^{\mu\lambda}g^{\nu\rho}\,^*\!H_{\lambda\rho}.  
\label{acons4}
\end{gather}
Writing out $^*\!H_{0i}$ using 
Eqs.\ (\ref{adual1})-(\ref{acons3}) 
one finds
\begin{equation}  \label{aHEB}
H_i=-\frac{\varepsilon_0c^2}{\sqrt{-g}}
\left(g_{00}g_{ij}-g_{i0}g_{j0}\right)B_j+
\frac{\varepsilon_0c}{\sqrt{-g}}[jkl]g_{j0}g_{ik}E_l \,.
\end{equation}
Comparison of this with Eqs.\ (\ref{aecon}) and (\ref{aDcons}) 
shows that
\begin{equation}
B_i=-\frac{\sqrt{-g}}{\varepsilon_0c^2g_{00}}g^{ij}H_j-
\frac{1}{\varepsilon_0cg_{00}}[ijk]g_{j0}E_k,
\end{equation}
which is the second of 
the constitutive equations 
(\ref{eq:constitution}) with (\ref{eq:constrel}). 

Clearly, several other relations between  
$\mathbf{D}$, $\mathbf{H}$, $\mathbf{E}$ and $\mathbf{B}$ 
are contained in (\ref{acons1})-(\ref{acons2}) and 
(\ref{acons3})-(\ref{acons4}). 
For example, to express $\mathbf{D}$ and $\mathbf{H}$ 
in terms of $\mathbf{E}$ and $\mathbf{B}$, 
as is done in Eq.\ (\ref{bhcons}), 
we need only take the time-space components 
of (\ref{acons1}), obtaining
\begin{equation}  \label{aDEB}
D_i=\varepsilon_0\sqrt{-g}\left(
g^{i0}g^{j0}-g^{ij}g^{00}\right)E_j-
\varepsilon_0c\sqrt{-g}[jkl]g^{k0}g^{ij}B_l \,,
\end{equation}
and the required formulae are 
Eqs.\ (\ref{aHEB}) and (\ref{aDEB}).

\renewcommand{\theequation}{B\arabic{equation}}
\setcounter{equation}{0}

\section*{Appendix B}
In this appendix we show that the expressions 
for $\varepsilon^{ij}$ and $\mu^{ij}$ 
derived in Ref.\ \cite{Ward} 
and utilized in Ref.\ \cite{PSS} 
are a special case of the formalism used in this paper. 
Consider the effective $\varepsilon^{ij}$ and $\mu^{ij}$ 
resulting from a transformation 
of the {\it spatial} coordinates of a Minkowski system. 
From Eq.\ (\ref{eq:constrel}) we obtain
\begin{equation} \label{aptpb}
\varepsilon^{ij}=-\sqrt{-g}g^{ij}\varepsilon'\,, 
\quad \mu^{ij}=-\sqrt{-g}g^{ij}\mu' \,,
\end{equation}
where we have allowed for general isotropic 
permittivity and permeability 
$\varepsilon'\neq1$, $\mu'\neq1$
in electromagnetic space. 

Rescale the spatial coordinate basis vectors $\partial_i$ 
so that they form a (in general non-coordinate) 
basis $\mathbf{u}_i$ of vectors of unit length:
\begin{equation} \label{audef}
\mathbf{u}_i=\frac{1}{\sqrt{-g_{ii}}}\,\partial_i 
\,, \quad \mathbf{g}(\mathbf{u}_i,\mathbf{u}_i)=1 \,. 
\qquad \text{(No summation.)}
\end{equation}
Let $\bar{g}_{ij}$ 
be the components of the metric tensor 
in the basis $\mathbf{u}_i$:
\begin{equation} \label{agbar}
\bar{g}_{ij}=
\mathbf{g}(\mathbf{u}_i,\mathbf{u}_j)=
\frac{1}{\sqrt{g_{ii}g_{jj}}}\,g_{ij} \,. \qquad 
\text{(No summation.)}
\end{equation}
We need to compute the triple product 
$\mathbf{u}_1\cdot(\mathbf{u}_2\times\mathbf{u}_3)$ 
in the coordinate basis  $\partial_i$; 
for this we require the 3D Levi-Civita tensor in this basis:
\begin{equation} \label{a3LC}
\varepsilon_{ijk}=\sqrt{-g}\,[ijk]\,, 
\quad \varepsilon^{ijk}=-\frac{1}{\sqrt{-g}}\,[ijk] \,,
\end{equation}
where $g$ is both the determinant of the space-time metric 
and the negative determinant of the spatial metric 
$g_{ij}$ since $g_{00}=1$, $g_{i0}=0$. 
Using Eqs.\ (\ref{audef})--(\ref{a3LC}) we find
\begin{eqnarray}
\mathbf{u}_1\cdot(\mathbf{u}_2\times\mathbf{u}_3)
&=&- g_{ij}u_1^i\varepsilon^{jkl}u_{2k}u_{3l}
=-g_{ij}u_1^i\varepsilon^{jkl}g_{km}u_2^mg_{ln}u_3^n 
\nonumber \\[8pt]
&=&-g_{1j}\frac{1}{\sqrt{-g_{11}}}\varepsilon^{jkl}g_{k2}
\frac{1}{\sqrt{-g_{22}}}g_{l3}\frac{1}{\sqrt{-g_{33}}}  
\nonumber \\[8pt]
&=&\frac{1}{\sqrt{-g_{11}g_{22}g_{33}}}
\frac{1}{\sqrt{-g}}[jkl]g_{1j}g_{2k}g_{3l}
=\frac{1}{\sqrt{-g_{11}g_{22}g_{33}}}
\frac{1}{\sqrt{-g}}g \nonumber \\[8pt]
&=&\frac{\sqrt{-g}}{\sqrt{-g_{11}g_{22}g_{33}}} \,. 
\label{atrip}
\end{eqnarray}
Using Eqs.\ (\ref{agbar}) and (\ref{atrip}) 
we write (\ref{aptpb}) as
\begin{gather}
\varepsilon^{ij}=
-\varepsilon'\mathbf{u}_1\cdot(\mathbf{u}_2\times\mathbf{u}_3)
\sqrt{-g_{11}g_{22}g_{33}}
\frac{1}{\sqrt{g_{ii}g_{jj}}}\bar{g}^{ij}\,,  \qquad 
\text{(No summation.)} \\[8pt]
\mu^{ij}=
-\mu'\mathbf{u}_1\cdot(\mathbf{u}_2\times\mathbf{u}_3)
\sqrt{-g_{11}g_{22}g_{33}}
\frac{1}{\sqrt{g_{ii}g_{jj}}}\bar{g}^{ij}\,,  \qquad 
\text{(No summation.)}
\end{gather}
which are the expressions derived in Ref.\ \cite{Ward}.
Note that this form of the theory implicitly contains 
the case of negative refraction,
because for a transition to a left-handed system 
the cross products change sign,
but this fact was never mentioned nor used so far.


\end{document}